%% file: main.tex
\documentclass[sigconf]{acmart}

\AtBeginDocument{%
  }

\usepackage{amsmath,amsthm,mathtools}
\usepackage{booktabs}
\usepackage{multirow}
\usepackage{listings}
\usepackage{xcolor}
\usepackage{wasysym}
\usepackage{threeparttable}
\usepackage{listings}
\usepackage{xcolor} 
\usepackage{graphicx}
\usepackage{subcaption}
\usepackage{enumitem}
\usepackage{balance}
\usepackage{tikz}
\usepackage[most]{tcolorbox}
\usepackage{tabularx}
\usepackage{array}

\usetikzlibrary{positioning,arrows.meta}

\begin{document}

\title{Debugging the Debuggers: Failure-Anchored Structured Recovery for Software Engineering Agents}


\author{Chenyu Zhao}
\affiliation{%
  \institution{Nankai University}
  \city{Tianjin}
  \country{China}}

\author{Shenglin Zhang}
\authornote{Corresponding author.}
\affiliation{%
  \institution{Nankai University}
  \city{Tianjin}
  \country{China}}

\author{Yihang Lin}
\affiliation{%
  \institution{Nankai University}
  \city{Tianjin}
  \country{China}}

\author{Wenwei Gu}
\affiliation{%
  \institution{Nankai University}
  \city{Tianjin}
  \country{China}}

\author{Zhimin Chen}
\author{Yongqian Sun}
\affiliation{%
  \institution{Nankai University}
  \city{Tianjin}
  \country{China}}

\author{Dan Pei}
\affiliation{%
  \institution{Tsinghua University}
  \city{Beijing}
  \country{China}}






\author{Chetan Bansal}
\author{Saravan Rajmohan}
\affiliation{%
  \institution{Microsoft}
  \city{Redmond}
  \country{USA}}

\author{Minghua Ma}
\affiliation{%
  \institution{Microsoft}
  \city{Redmond}
  \country{USA}}







\renewcommand{\shortauthors}{Zhao et al.}


\newcommand{\ie}{\textit{i.e.,}~}
\newcommand{\eg}{\textit{e.g.,}~}
\newcommand{\etal}{\textit{et al.}~}
\def\name{\textit{PROBE}}

\newcommand{\best}[1]{\textbf{#1}}

\begin{abstract}

\input{sections/abstract}

\end{abstract}

\ccsdesc[500]{Software and its engineering~Software maintenance tools}
\ccsdesc[300]{Computing methodologies~Planning and scheduling}
\ccsdesc[100]{Computer systems organization~Maintainability and maintenance}
\ccsdesc[100]{Computing methodologies~Knowledge representation and reasoning}

\keywords{LLM agents, AgentOps, failure recovery, telemetry, structured diagnosis, bounded guidance}


\maketitle

\section{Introduction}
\input{sections/intro}

\section{Background and Motivation}
\label{sec: background}

\input{sections/background}

\input{sections/method}

\input{sections/experiment}

\input{sections/deployment}

\input{sections/discussion}

\section{Conclusion}

\input{sections/conclusion}

\section{Data Availability}
\input{sections/data_availablity}

\bibliographystyle{ACM-Reference-Format}
\bibliography{main}










\end{document}

%% file: sections/abstract.tex
Software engineering agents are increasingly deployed in evaluable engineering environments, yet post-failure recovery remains costly, manual, and largely ad hoc.
Existing systems expose traces for inspection or generate follow-up feedback, but they do not turn heterogeneous runtime evidence into grounded, bounded recovery guidance for a subsequent attempt.
We present \name{}, a failure-anchored framework for structured recovery in software engineering agents.
\name{} organizes failed-run telemetry into structured evidence, structured diagnosis, and bounded recovery guidance, realized by a Telemetry Layer, a Diagnosis Layer, and a Guidance Gate.
The Telemetry Layer preserves fine-grained runtime signals, the Diagnosis Layer fuses cross-signal evidence into a grounded diagnosis, and the Guidance Gate constructs diagnosis-derived recovery guidance only when it is evidence-grounded, actionable, and within the scope of agent-side behavior.

We evaluate \name{} across three software engineering settings spanning repository-level software repair, enterprise workflow recovery, and AIOps service mitigation.
On 257 initially unresolved cases, \name{} achieves 65.37\% Top-1 diagnosis accuracy and a 21.79\% recovery rate, outperforming the strongest non-\name{} baseline by 43.58 and 12.45 percentage points, respectively.
The results reveal a substantial diagnosis--recovery gap: accurate diagnosis is necessary but not sufficient unless it can be translated into bounded guidance that the subsequent attempt can execute and verify.
Beyond controlled evaluation, a Microsoft IcM prototype shows that \name{} can attach as a non-intrusive side channel to existing service-diagnosis agent workflows without changing the agent policy, toolset, or execution budget.
These results suggest that telemetry-grounded, failure-anchored recovery is a practical path to improving post-failure recoverability of software engineering agents under realistic engineering constraints.

%% file: sections/intro.tex
\begin{figure}[t]
\centering
\includegraphics[width=0.9\linewidth]{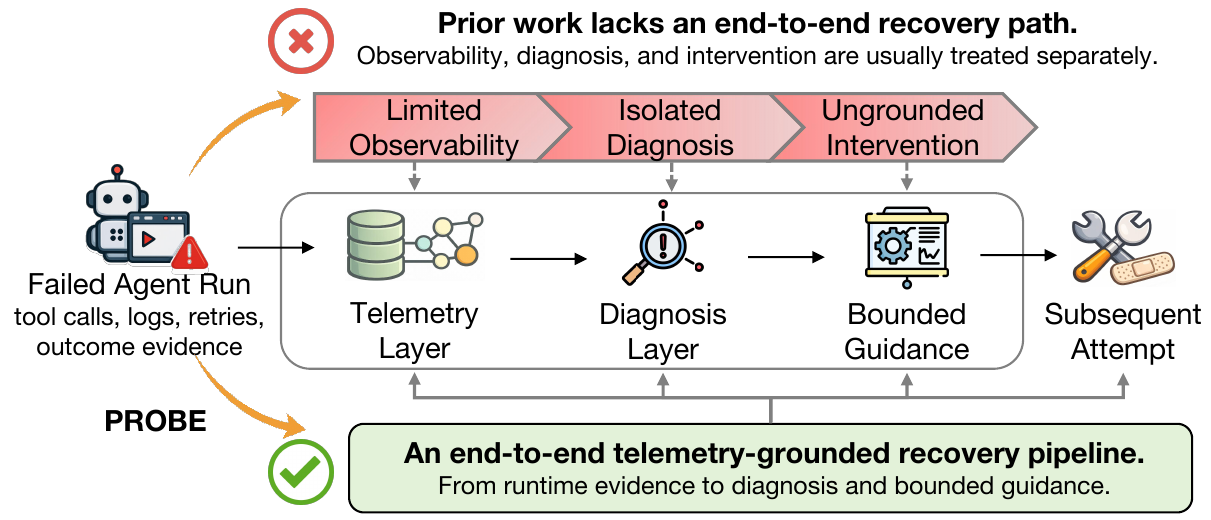}
\caption{Prior work treats observability, diagnosis, and intervention separately. \name{} connects runtime evidence, diagnosis, and bounded guidance into one recovery loop.}
\label{fig:overview}
\end{figure}

Software engineering agents powered by large language models are increasingly deployed in complex real-world settings, including repository-level code repair~\citep{bairi2024codeplan,yang2024sweagent,xia2025agentless}, enterprise workflow automation~\citep{malay2026enterpriseopsgym}, and cloud service operations~\citep{DBLP:conf/mlsys/ChenSSMSMBWR25,DBLP:conf/icse/AhmedGBZZR23,chen2024rcacopilot,jiang2024xpert}.
As these agents tackle broader and more complex tasks, failed executions have become a pervasive practical concern~\cite{xia2025agentless,olausson2024selfrepair}.
In practice, such failures are often handled by inspecting a few error messages, appending prompt instructions, and rerunning the agent~\cite{xia2024chatrepair,2024RepairAgent}.
However, this ad hoc approach is unreliable: naive retry has been shown to yield diminishing returns without targeted feedback~\cite{olausson2024selfrepair}.
More importantly, such recovery is weakly tied to failed-run provenance, obscuring which evidence should guide the subsequent attempt and which behaviors should be avoided.

In practice, failed executions produce rich runtime evidence.
However, most existing benchmarks focus primarily on whether the task objective is satisfied and do not preserve the fine-grained diagnostic signals needed to support recovery~\cite{jimenez2024swebench,malay2026enterpriseopsgym}.
Recent analyses further show that many agent failures stem from recurring behavioral patterns rather than random errors~\citep{bouzenia2025trajectories,nanda2026,xia2025agentless}.
Process-oriented studies show that execution errors can be categorized at the step level and that recurring coding-agent misbehaviors, such as specification drift, hallucinated reasoning, and malformed tool calls, can often be addressed through targeted guidance~\cite{chen2026beyondfinalcode,nanda2026}.
These findings indicate that failed executions already contain actionable evidence for subsequent attempts, yet existing mechanisms do not systematically turn such evidence into diagnosis and recovery guidance.

Recent work and emerging practice have begun to make agent executions more observable, measurable, and diagnosable~\cite{GoogleAgentsCompanion2025}.
However, as Figure~\ref{fig:overview} illustrates, existing efforts remain fragmented: observability systems expose telemetry and traces for inspection~\cite{dong2024agentops,langsmith2026docs}; diagnosis-oriented methods localize critical failure steps or generate explanations~\cite{chen2023selfdebug,barke2026agentrx,tan2026llmrca}; and iterative refinement methods provide feedback for subsequent attempts~\cite{shinn2023reflexion,madaan2023selfrefine}, but such feedback is typically weakly grounded in the runtime telemetry most relevant to recovery.
Taken together, prior work improves inspection, explanation, or retry feedback, but does not connect runtime telemetry, failure interpretation, and follow-up intervention into a single recovery process.

We make this missing recovery process concrete as a progression from telemetry to evidence, diagnosis, and guidance.
Runtime telemetry can be organized into \emph{structured evidence}: localized and fused failure-relevant signals preserved with provenance.
Such evidence can support a \emph{structured diagnosis}: a schema-constrained account of the failure anchor, primary cause, behavioral mistake, supporting evidence, and confidence.
The diagnosis can then be translated into \emph{bounded recovery guidance}: actionable and verifiable instructions for the subsequent attempt.
This progression leads to three practical challenges.

\textbf{Challenge 1. Preserving recovery-critical evidence.}
Effective recovery depends on more than a final error message or a coarse execution summary.
It requires retaining evidence such as exception signatures, repeated tool failures, execution order, agent and environment state, and evaluator feedback.
Since this evidence is scattered across execution steps, it is easily lost when telemetry is compressed into generic summaries or undifferentiated logs before being organized into structured evidence.

\textbf{Challenge 2. Building structured diagnosis.}
Runtime signals differ in source, format, granularity, and reliability.
Metrics capture progress and resource use, logs record localized failures, traces preserve temporal structure, while status, environment, and outcome signals provide contextual information at the execution and task levels.
A useful framework should integrate these signals without collapsing their distinct roles, and produce a structured diagnosis that remains traceable to evidence and usable for recovery guidance.

\textbf{Challenge 3. Producing grounded and bounded guidance.}
A diagnosis explains the failure, but not how the subsequent attempt should act.
Guidance must be evidence-grounded to avoid unsupported recovery instructions, and bounded to constrain the agent to verifiable, in-scope actions.
A practical system must therefore translate diagnosis into guidance with an explicit target, operation, verification signal, and boundary condition.

To address these challenges, we present \name{}, an end-to-end framework for failure-anchored structured recovery in software engineering agents.
\name{} uses failed execution telemetry as the anchor for recovery and realizes the progression from structured evidence to structured diagnosis and bounded recovery guidance through a \emph{Telemetry Layer}, a \emph{Diagnosis Layer}, and a \emph{Guidance Gate}.
The \emph{Telemetry Layer} instruments the agent run at span level and organizes failed-run telemetry into typed signal families, including metrics, logs, traces, agent intent, tool--environment state, and optional external outcome signals.
The \emph{Diagnosis Layer} localizes failure-relevant signals and fuses them into structured evidence, then derives a structured diagnosis through anchor-first diagnosis generation.
The \emph{Guidance Gate} acts as a grounding, actionability, and scope filter, converting the diagnosis into bounded recovery guidance only when it is telemetry-supported, actionable, and within the scope of agent-side behavior.

To make the role of the \emph{Guidance Gate} concrete, consider a service-remediation case from AIOpsLab (Section~\ref{sec:eval-case-study}).
A structured diagnosis can identify a misconfigured \texttt{targetPort} and premature submission, but it does not determine how the subsequent attempt should proceed.
\name{} therefore translates the diagnosis into bounded recovery guidance by specifying the target to revisit, the operation to perform, the verification signal to check, and the boundary condition that prevents premature completion.


Our contributions are as follows.

\begin{itemize}[leftmargin=*]
    \item We formulate \emph{failure-anchored structured recovery} for software engineering agents as a transformation from failed-run telemetry to structured evidence, structured diagnosis, and bounded recovery guidance for a subsequent attempt.


    \item We implement \name{} as a portable, framework-agnostic Python package that provides side-channel recovery support through span-level telemetry, evidence-aware diagnosis, and generation of grounded and bounded guidance, without modifying the agent policy, toolset, executor, or evaluator.
    
    \item We evaluate \name{} across three software engineering settings: SWE-bench~\cite{jimenez2024swebench}, EnterpriseOps-Gym~\cite{malay2026enterpriseopsgym}, and AIOpsLab~\cite{DBLP:conf/mlsys/ChenSSMSMBWR25}. 
    On 257 initially unresolved cases, \name{} achieves 65.37\% Top-1 diagnosis accuracy and a 21.79\% recovery rate, outperforming the strongest baseline by 43.58 and 12.45 percentage points while revealing a diagnosis--recovery gap.

    \item We validate the industrial integration boundary through a non-intrusive Microsoft IcM service-diagnosis prototype, showing that \name{} can attach to existing service-diagnosis agent workflows without modifying the agent policy, toolset, or execution budget.
\end{itemize}

%% file: sections/background.tex
\begin{table*}[t]
\centering
\small
\setlength{\tabcolsep}{3.8pt}
\caption{Human-annotated failure-cause taxonomy for the 257 initially unresolved cases.}
\label{tab:motivation-failure-taxonomy}
\begin{tabular*}{\textwidth}{@{\extracolsep{\fill}}p{0.21\textwidth}p{0.3\textwidth}p{0.33\textwidth}c@{}}
\toprule
Reviewed failure cause & Keywords & Explanation & Count \\
\midrule
Insufficient validation & missing verification, outcome unchecked, incomplete confirmation & The agent proceeded without adequately verifying whether the intended outcome was achieved. & 83 \\
Tool/subprocess failure handling & tool anomaly, timeout, execution failure, invalid output & The agent failed to correctly recognize or respond to tool-level execution anomalies. & 47 \\
State/workflow error & wrong target, state mismatch, missing field, workflow violation & The agent operated on the wrong target or left the workflow in an incorrect final state. & 42 \\
Patch/submission workflow & premature submission, incomplete artifact, workflow breakdown & The run broke down near artifact production or submission, leaving the final deliverable incomplete or invalid. & 35 \\
Retry/no-progress loop & repeated retries, no progress, no adaptation, stalled loop & The agent repeated ineffective actions without making meaningful progress or adapting strategy. & 26 \\
Runtime/environment handling & environment precondition, missing dependency, infrastructure blocker & Execution was blocked by an unresolved environment or infrastructure precondition. & 24 \\
\bottomrule
\end{tabular*}
\end{table*}

\subsection{Agent Stacks and Execution Layers}

Modern software engineering agents are typically realized as layered, tool-using systems rather than single model calls.
Across code repair, enterprise workflow automation, and AIOps service mitigation, runs involve multi-step interactions with tools and execution environments~\cite{yang2024sweagent,xia2025agentless,bairi2024codeplan,barke2026agentrx}.
These agents expose recurring execution boundaries, including model calls, orchestration decisions, tool interactions, workflow state updates, and evaluator feedback~\citep{MCPagentbench,zhou2024webarena,trivedi2024appworld,wu2024autogen}.
Representative stacks such as LiteLLM~\cite{berriai2024litellm}, LangChain-style orchestration~\cite{chase2022langchain}, MCP~\cite{mcp2025spec}, and tool/API-oriented agents~\citep{schick2023toolformer,qin2024toolllm,patil2024gorilla,li2023apibank} instantiate these boundaries in different ways.

This common boundary structure motivates \name{}'s framework-agnostic design.
Rather than encoding recovery logic for a specific agent framework, \name{} records recovery-relevant signals through a shared telemetry schema that can be instantiated across heterogeneous stacks~\citep{zhou2024webarena,trivedi2024appworld,wu2024autogen,xie2024openagents}.
In our implementation, the same schema supports LiteLLM-based SWE-agent runs~\cite{yang2024sweagent}, LangChain-based ReAct with MCP-mediated tool access in EnterpriseOps-Gym~\cite{malay2026enterpriseopsgym}, and AIOpsLab~\cite{DBLP:conf/mlsys/ChenSSMSMBWR25}, keeping the recovery pipeline unchanged across settings.

\subsection{Related Work}

Prior work covers several ingredients of recovery for software engineering agents, but usually treats them as separate problems rather than as stages of a single recovery process.
Observability systems such as AgentOps~\cite{dong2024agentops} and LangSmith~\cite{langsmith2026docs} collect telemetry and expose execution traces for inspection, but they primarily treat telemetry as an inspection artifact rather than converting it into structured diagnosis or bounded recovery guidance.
Diagnosis-oriented work studies failure localization, explanation, and debugging support~\citep{barke2026agentrx,bouzenia2025trajectories,chen2023selfdebug,chen2024rcacopilot,jiang2024xpert,xie2024latentscope};
for example, AgentRx~\citep{barke2026agentrx} localizes critical failure steps with a cross-domain taxonomy, and LLMRCA~\citep{tan2026llmrca} shows that metrics, logs, and traces can support multilevel diagnosis for deployed LLM applications.
Cloud-incident systems further show the value of using historical incidents, runtime diagnostic information, and domain-specific queries for diagnosis and incident investigation~\citep{DBLP:conf/icse/AhmedGBZZR23,chen2024rcacopilot,jiang2024xpert,zhang2024gpt4rca}.
Traceability-oriented work such as CodeTracer~\citep{li2026codetracer} reconstructs code-agent logs into hierarchical trace trees, localizes failure-responsible stages, and uses localized diagnostic signals for reflective replay.
Refinement methods such as Reflexion~\cite{shinn2023reflexion} and Self-Refine~\cite{madaan2023selfrefine} generate feedback for a later attempt, but the feedback is often only weakly grounded in runtime telemetry and is typically not bounded by a structured diagnosis.
Overall, existing methods either expose traces for inspection, localize failure-relevant steps for explanation or replay, or provide generic feedback for retry.
What remains missing is a recovery process that treats failed-run telemetry as the anchor, converts multi-signal evidence into a structured diagnosis, and admits only grounded, actionable, and in-scope guidance into the subsequent attempt.

\subsection{Motivation}
\label{motivation}
We focus on software engineering agents whose executions admit task-level evaluation, where each run can be judged as success or failure by an environment-specific evaluator, verifier, or benchmark harness.
To understand what makes failed executions recoverable, we analyze the 257 initially unresolved first-attempt reports used in our recovery evaluation: 102 from SWE-bench~\cite{jimenez2024swebench}, 106 from EnterpriseOps-Gym~\cite{malay2026enterpriseopsgym}, and 49 from AIOpsLab~\cite{DBLP:conf/mlsys/ChenSSMSMBWR25}.

For each report, two professional software engineers independently inspect the saved execution trajectory, runtime records, and outcome evidence, without access to \name{}'s structured diagnosis or gated guidance, and assign a reviewed failure-cause category.
To avoid circular evaluation, the annotators do not use \name{}'s structured diagnosis or gated guidance when assigning the reviewed failure cause.
Disagreements are resolved through discussion and category consolidation.
This analysis clarifies what recovery-relevant evidence is already present in failed executions before any recovery feedback is generated, and why final outcome labels alone are insufficient for iterative recovery.

\subsubsection{Finding 1. Initially unresolved runs are dominated by process-level failures}

The initially unresolved reports are not dominated by a single surface error.
Instead, they exhibit recurring process-level failure modes, as summarized in Table~\ref{tab:motivation-failure-taxonomy}.
Across all 257 reports, the largest categories are insufficient validation, tool/subprocess failure handling, and state/workflow error, which together account for 172 cases (66.93\%).

This distribution suggests that failed executions are better understood as process failures than as isolated final verdicts.
Many failures arise from how the agent interprets tool feedback, verifies intermediate results, reacts to runtime conditions, and decides whether a run is ready to terminate or submit.
These results motivate recovery support that operates on execution evidence rather than relying only on a final failure signal.

\subsubsection{Finding 2. Effective recovery requires heterogeneous runtime signals}

The failure categories summarized in Table~\ref{tab:motivation-failure-taxonomy} cannot be reliably identified from final evaluator outcomes alone.
A final unresolved verdict indicates that the task objective was not satisfied, but it rarely explains whether the failure comes from missing validation, broken workflow execution, repeated tool errors, incorrect state transitions, or runtime/environment handling.
These distinctions require inspecting how the execution unfolded over time.


Different failure categories depend on different combinations of runtime evidence.
For example, insufficient validation requires linking intermediate actions, evaluator outputs, and the agent's decision to stop or submit.
Tool/subprocess failure handling depends on tool-call sequences, return codes, and whether the agent adapts after failure.
Runtime/environment-related failures require distinguishing agent-recoverable configuration issues from external infrastructure blockers.
No single raw signal is sufficient across these categories.

These observations motivate the design of \name{}.
To support iterative recovery for software engineering agents, the system should preserve fine-grained runtime records, organize heterogeneous signals into structured evidence, diagnose the actionable failure cause, and pass only bounded, evidence-grounded guidance to the subsequent attempt.

%% file: sections/method.tex
\begin{figure*}[!t]
\centering
\includegraphics[width=0.8\textwidth]{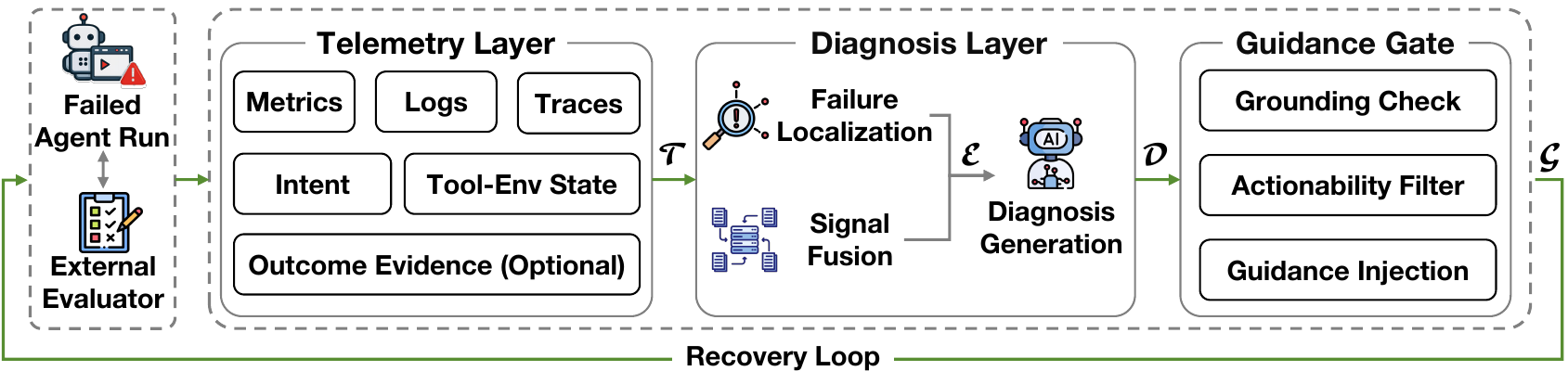}
\caption{Overview of the \name{} framework. \name{} organizes failed-run telemetry into a typed telemetry bundle \((\mathcal{T})\), constructs structured evidence \((\mathcal{E})\), derives structured diagnosis \((\mathcal{D})\), and produces bounded recovery guidance \((\mathcal{G})\), forming a recovery loop for subsequent attempts.}
\label{fig:probe_framework}
\end{figure*}

\section{PROBE}
\label{sec:probe}

\subsection{Overview}
\label{sec:probe:overview}

\name{} models post-failure recovery for software engineering agents as a structured transformation from failed-run telemetry to bounded recovery guidance for the subsequent attempt.
As shown in Figure~\ref{fig:probe_framework}, the \emph{Telemetry Layer} records failed-run telemetry into typed signal families, the \emph{Diagnosis Layer} localizes and fuses these signals into structured evidence before deriving a structured diagnosis, and the \emph{Guidance Gate} validates and constructs bounded recovery guidance.
Formally, let a failed run produce a telemetry bundle
\begin{equation}
\label{eq:telemetry-bundle}
\mathcal{T} =
\{T_{\text{metrics}}, T_{\text{logs}}, T_{\text{traces}}, T_{\text{intent}}, T_{\text{env}}, T_{\text{outcome}}\},
\end{equation}
where each component corresponds to one signal family, and \(T_{\text{outcome}}\) is included when external evaluator signals are available.

\name{} processes the telemetry bundle through the structured pipeline
\begin{equation}
\label{eq:recovery-pipeline}
\mathcal{T}
\rightarrow
\mathcal{E}
\rightarrow
\mathcal{D}
\rightarrow
\mathcal{G},
\end{equation}
where \(\mathcal{T}\) denotes the typed telemetry bundle, \(\mathcal{E}\) denotes structured evidence formed by localizing and fusing failure-relevant telemetry signals, \(\mathcal{D}\) denotes the structured diagnosis object, and \(\mathcal{G}\) denotes bounded recovery guidance.
This benchmark-agnostic schema abstracts model-access, orchestration, and tool-interface events into a shared telemetry representation, enabling \name{} to instrument existing agents without changing their policy, toolset, executor, or evaluator.

\subsection{Telemetry Layer}
\label{sec:probe:telemetry}
To preserve recovery-critical evidence, the \emph{Telemetry Layer} records runtime events as span-level telemetry and organizes them into typed signal families.
The resulting telemetry bundle \(\mathcal{T}\) preserves signal provenance for downstream evidence construction and diagnosis.

\subsubsection{Telemetry Representation}
\label{sec:probe:telemetry-representation}

The \emph{Telemetry Layer} represents a software engineering agent run using four notions: spans, traces, intent, and tool--environment state.
A \emph{span} \(\sigma_t\) is the smallest recorded execution unit at step \(t\), such as a model response, tool call, verifier result, metric snapshot, or runtime exception.
Each span records its event type, timestamp, payload, status, and available tool, model, or error metadata.

A \emph{trace} is the temporally ordered sequence of spans from one run.
It preserves execution structure such as repeated cycles and delayed verification.
To capture agent-side behavior, \name{} augments each step with intent \(i_t\), which describes what the agent is trying to do, such as gathering evidence, editing artifacts, running verification, or preparing submission.
When explicit phase labels are unavailable, \name{} infers intent from model messages, tool choices, command types, verifier calls, and submission events.
To capture environment-side context, \name{} records tool--environment state \(s_t\), including tool-return status, workspace or workflow state, and evaluator-side signals.
Together, \(i_t\) and \(s_t\) distinguish what the agent intended to do from what the environment actually exposed or changed.

\subsubsection{Signal Families}
\label{sec:probe:signal-families}

Given this representation, \name{} partitions collected telemetry into signal families with distinct diagnostic roles.


\textbf{Metrics.}
\name{} derives \(T_{\text{metrics}}\) from runtime counters, span-level metadata, and scalar features computed from intent \(i_t\) and tool--environment state \(s_t\).
We organize these metrics into four recovery-oriented groups.
\emph{Cost and capacity} capture token velocity, context saturation, tool-call density, and retry dominance.
\emph{Recovery progress} measures observable state changes in \(s_t\), or uses intent transitions in \(i_t\) as a fallback when explicit workflow progress is unavailable.
\emph{Progress--cost coupling} characterizes whether observed progress is commensurate with the resources consumed in the same window.
\emph{Behavioral stability} captures intent volatility, intent run-length ratio, and tool-switch volatility.
Together, these groups help identify runs that make little progress, consume resources without corresponding state change, or fall into repetitive or unstable action patterns.

\textbf{Logs.}
\(T_{\text{logs}}\) is extracted from tool returns, command outputs, parser failures, runtime exceptions, and system messages.
\name{} deduplicates repeated messages, truncates long outputs, and canonicalizes common errors into reusable signatures.
This preserves local failure anchors such as repeated errors, invalid tool arguments, subprocess failures, and environment-side breakdowns.

\textbf{Traces and intent.}
\(T_{\text{traces}}\) and \(T_{\text{intent}}\) capture the temporal structure and functional role of each step in the run.
They expose repeated failures, missing progress, unstable action transitions, and role shifts across the run.

\textbf{Tool--environment state.}
\(T_{\text{env}}\) records environment-side context, including tool availability, tool-return status, workspace or workflow state, and evaluator-side observations.
It helps distinguish agent-side behavioral failures from environment-side constraints.

\textbf{Optional external outcome evidence.}
\(T_{\text{outcome}}\) is included when evaluator-side feedback is available, such as benchmark verdicts, test results, incident-resolution checks, or task-specific scoring scripts.
It grounds telemetry against the task objective and helps detect mismatches between agent claims and evaluator observations.



\subsection{Diagnosis Layer}
\label{sec:probe:diagnosis}

\begin{figure}[t]
\centering
\includegraphics[width=\linewidth]{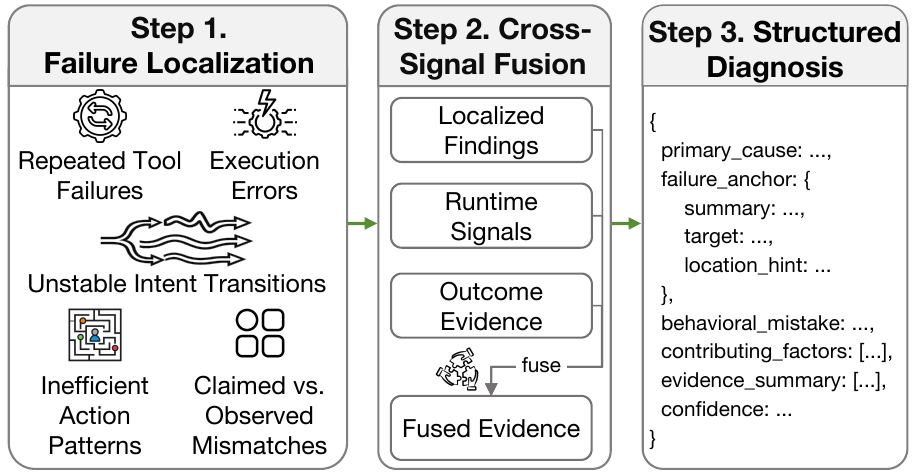}
\caption{Detailed workflow of the \emph{Diagnosis Layer}.}
\label{fig:diagnosis-layer}
\end{figure}

To support bounded recovery guidance, the \emph{Diagnosis Layer} converts \(\mathcal{T}\) into structured evidence \(\mathcal{E}\) and then into structured diagnosis \(\mathcal{D}\) through failure localization, cross-signal fusion, and structured diagnosis generation, as shown in Figure~\ref{fig:diagnosis-layer}.

\subsubsection{Failure Localization}
\label{sec:probe:failure-localization}


Given telemetry bundle \(\mathcal{T}\), failure localization applies signal-specific detectors to identify failure-relevant steps, events, and execution windows as typed localized findings.

For metrics, \name{} uses robust window-based anomaly scoring: it computes robust z-scores with the median absolute deviation (MAD)~\citep{ruff2021unifying}, derives tail thresholds from empirical quantiles, and emits findings when metrics cross these thresholds.
When no metric dominates, \name{} applies Isolation Forest~\citep{han2022adbench} to window-level feature vectors and retains the highest-scoring window as aggregate metric evidence.

For logs and tool--environment state, \name{} localizes execution errors and repeated failures from tool returns, verification records, subprocess returns, guardrail signals, and error-bearing log segments.
Related evidence is grouped by stable anchors such as tool name, argument fingerprint, error signature, return code, and nearby control context.
When an auxiliary model is available, \name{} derives grounded semantic error signatures; otherwise, it retains raw signatures and infrastructure clues such as timeout, connection, or out-of-memory signals.

For traces and intent, \name{} localizes unstable intent transitions and run-level action patterns from the ordered execution history.
It scores local transitions with a smoothed \(n\)-gram model and emits findings for transitions in the high-surprise tail.
For longer-range patterns, it groups repeated failed actions using stable keys and summarizes the run timeline together with recent verification records, outcome evidence, and metric findings.

When external outcome evidence is available, evaluator verdicts and failing checks are converted into typed findings so that internal execution evidence can be compared with the task objective.

\subsubsection{Cross-Signal Fusion}
\label{sec:probe:signal-fusion}

Cross-signal fusion aligns localized findings that refer to the same failure-relevant event, behavior pattern, or outcome condition.
It converts localized findings into fused evidence records, where each record preserves a shared anchor, supporting signal sources, temporal scope, severity, and agreement or conflict across signals.

To construct fused evidence, \name{} first normalizes each localized finding into a standard evidence unit containing an anchor, source, time scope, severity, and evidence reference.
It then groups compatible units by anchor, time, or semantic content and aggregates each group into a fused evidence record that preserves both support and disagreement.
For example, repeated failures from logs, traces, and outcome evidence can strengthen the same unresolved-failure anchor, while an internal success claim contradicted by evaluator feedback is preserved as a conflict.


The resulting fused evidence records form structured evidence \(\mathcal{E}\), preserving anchors, support, conflicts, and signal provenance for downstream structured diagnosis.

\subsubsection{Structured Diagnosis}
\label{sec:probe:structured-rca}

Given fused evidence \(\mathcal{E}\), \name{} constructs a structured diagnosis through an anchor-first diagnosis protocol.
Following decomposition-based prompting methods~\citep{wei2022chain,zhou2023least,yao2023react}, the default implementation instantiates this protocol with an LLM-based diagnosis model prompted to emit a schema-constrained diagnosis object.
When the model is unavailable, \name{} falls back to a deterministic summarizer built from localized findings.
The goal is not to generate a free-form explanation, but to produce a run-specific diagnosis that is grounded in evidence and structured enough to support bounded recovery guidance.

First, the protocol is anchor-first.
Each diagnosis must start from a concrete, telemetry-supported failure anchor before making causal claims.
The anchor identifies where the failure becomes diagnostically actionable, such as a failed check, contradicted outcome, repeated execution error, inconsistent workflow state, or supported artifact/log location.
Unlike the reviewed failure-cause category, which describes the type of failure, the failure anchor grounds the diagnosis in the specific run.

Second, the protocol records guidance-relevant fields by distinguishing the \emph{primary cause} from the \emph{behavioral mistake}.
The primary cause explains why the task failed, while the behavioral mistake explains what the agent did or failed to do after the failure anchor appeared.
This distinction matters because the same failure anchor may require different recovery guidance, such as completing a missing action, avoiding repeated invalid tool use, delaying submission until verification succeeds, or revising an incorrect hypothesis.
Evidence that contextualizes but does not directly explain the main failure is recorded as \emph{contributing factors}.

Third, the protocol is schema-constrained and confidence-aware.
The structured diagnosis is emitted as a fixed object containing the \emph{primary cause}, \emph{failure anchor}, \emph{behavioral mistake}, \emph{contributing factors}, \emph{evidence summary}, and \emph{confidence}, matching the schema shown in Figure~\ref{fig:diagnosis-layer}.
The confidence field is a bounded estimate of evidential support for the diagnosis object, not a calibrated probability of correctness~\citep{zhang2024pace,zhang2024lmpace}.
\name{} clips model-emitted confidence to \([0,1]\), assigns conservative fallback confidence when the model is unavailable, limits contributing factors and evidence items, and requires failure-anchor fields to be supported by fused evidence.
Confidence alone does not authorize recovery guidance; the \emph{Guidance Gate} still applies grounding, actionability, and intervention-scope checks before the diagnosis can shape the subsequent attempt.


Together, these constraints make \(\mathcal{D}\) an evidence-grounded diagnosis object that can be checked before being converted into bounded recovery guidance.

\subsection{Guidance Gate}
\label{sec:probe:guidance-gate}
To turn structured diagnosis into guidance, the \emph{Guidance Gate} acts as a deterministic validation and injection layer between \(\mathcal{D}\) and the subsequent attempt.
Given a structured diagnosis \(\mathcal{D}\), the gate determines whether it can contribute \emph{diagnosis-derived recovery guidance} and constructs bounded recovery guidance \(\mathcal{G}\) only when the diagnosis is evidence-grounded, actionable, and within the agent's intervention scope.

\textbf{Grounding Check.}
To prevent unsupported assumptions from being promoted into recovery instructions, the grounding check verifies whether the diagnosis is anchored in telemetry-supported evidence.
A diagnosis is considered grounded only if its primary cause or failure anchor can be traced back to structured evidence, such as a failed evaluator check, repeated tool error, contradicted outcome, inconsistent workflow state, or concrete artifact/log location.
If an artifact path, entity name, service name, location hint, or failure location is not supported by fused evidence, the gate does not use it as a guidance target.
When the diagnosis lacks both a supported primary cause and a supported failure anchor, the diagnosis-derived guidance component is marked as non-injectable.

\textbf{Actionability Filter.}
To ensure that grounded diagnosis can guide concrete agent behavior, the actionability filter checks whether it can be operationalized as bounded recovery guidance.
Inspired by prior work on iterative feedback and self-refinement~\citep{shinn2023reflexion,madaan2023selfrefine}, operationalizable guidance must specify four elements: a target, an operation, a verification signal, and a boundary condition.
The target identifies what the agent should revisit; the operation specifies what it should do; the verification signal specifies what evidence should be checked afterward; and the boundary condition specifies when the agent should stop, avoid repetition, or return to evidence collection.
The filter also checks whether the proposed intervention falls within the scope of agent-side recovery.
Runtime and environment-related failures are not uniformly excluded, since some correspond to recoverable configuration, dependency, or validation issues.
However, diagnoses dominated by platform outage, persistent connection failure, out-of-memory error, Docker failure, container crash, or other infrastructure-level conditions are not converted into direct corrective guidance.
For such cases, the gate bounds the feedback around verification, evidence collection, or avoiding premature completion instead of issuing speculative corrective instructions.

\textbf{Guidance Injection.}
After the \emph{Grounding Check} and \emph{Actionability Filter}, the gate constructs bounded recovery guidance \(\mathcal{G}\) from the validated fields.
\(\mathcal{G}\) records whether diagnosis-derived guidance is injectable and, when injectable, specifies the target, operation, verification signal, and boundary condition for the subsequent attempt.
The final \emph{recovery hint block} is produced by a formatter rather than copied directly from \(\mathcal{G}\).
Injectable guidance is used as the main hint.
Non-injectable guidance is not converted into direct corrective instructions; instead, the formatter may add conservative telemetry-supported hints, such as re-checking evidence, rerunning verification, or avoiding premature submission.

Overall, the \emph{Guidance Gate} preserves the distinction between diagnosis and guidance by allowing only evidence-grounded, actionable, and bounded guidance to shape the subsequent attempt.

%% file: sections/experiment.tex
\begin{table*}[!t]
\centering
\caption{Recovery performance and diagnosis quality on initially unresolved cases. 
\emph{Rate} is the fraction of \emph{Initial Unres.} cases recovered after one repair attempt. 
\emph{Top-1} is the fraction of cases whose diagnosed failure cause matches the independently reviewed failure cause.}
\label{tab:main-recovery}
\small
\setlength{\tabcolsep}{3pt}
\begin{tabular*}{\textwidth}{@{\extracolsep{\fill}}lw{c}{1.05cm}w{c}{0.75cm}w{c}{0.75cm}w{c}{0.95cm}w{c}{0.75cm}w{c}{0.95cm}w{c}{0.75cm}w{c}{0.95cm}@{}}
\toprule
& & Outcome Only
& \multicolumn{2}{c}{AgentOps Summary}
& \multicolumn{2}{c}{LangSmith Summary}
& \multicolumn{2}{c}{\name{} Full Pipeline} \\
\cmidrule(lr){3-3}
\cmidrule(lr){4-5}
\cmidrule(lr){6-7}
\cmidrule(lr){8-9}
Setting & \shortstack{Initial Unres.}
& Rate
& Rate & Top-1
& Rate & Top-1
& Rate & Top-1 \\
\midrule
SWE-bench & 102
& 1.96\%
& 12.74\% & 12.75\%
& 14.71\% & 29.41\%
& \textbf{28.43\%} & \textbf{83.33\%} \\
EnterpriseOps-Gym & 106
& 0.00\%
& 5.66\% & 3.77\%
& 2.83\% & 3.77\%
& \textbf{11.32\%} & \textbf{57.55\%} \\
AIOpsLab & 49
& 8.16\%
& 10.20\% & 42.86\%
& 16.33\% & \textbf{44.90\%}
& \textbf{30.61\%} & \textbf{44.90\%} \\
\midrule
\textbf{Total} & 257
& 2.33\%
& 9.34\% & 14.79\%
& 9.34\% & 21.79\%
& \textbf{21.79\%} & \textbf{65.37\%} \\
\bottomrule
\end{tabular*}
\end{table*}

\section{Experiments}
\label{sec: experiment}

Our evaluation asks whether failed-run telemetry can be converted into useful recovery feedback for software engineering agents.
We study this question along three research questions:
\begin{itemize}[leftmargin=*]
    \item \textbf{RQ1: Recovery effectiveness.}
    Can \name{} improve diagnosis quality and recovery on initially unresolved runs compared with outcome-only feedback and observability-summary baselines?
    \item \textbf{RQ2: Actionable and gated recovery guidance.}
    Can \name{} transform failed-run evidence into actionable and bounded feedback for the subsequent attempt?
    \item \textbf{RQ3: Practical overhead.}
    How much telemetry does \name{} collect, and what token and latency overhead does it introduce when generating recovery feedback for the subsequent attempt?
\end{itemize}

\subsection{Experimental Setup}
\label{sec:eval-setup}

To evaluate post-failure recovery across heterogeneous software engineering settings, we use three benchmarks with explicit task-level evaluators.
Together, they cover repository-level repair, enterprise workflow execution, and AIOps service mitigation.

\subsubsection{Benchmarks and Recovery Protocol}
\label{sec:eval-benchmarks-protocol}

For repository-level software repair, we use SWE-bench~\cite{jimenez2024swebench} with mini-SWE-agent, where the agent must inspect a repository, edit source code, and produce a patch that satisfies the benchmark evaluator.
For enterprise workflow recovery, we use EnterpriseOps-Gym~\cite{malay2026enterpriseopsgym} with its default task executor.
For AIOps service mitigation, we use Microsoft AIOpsLab~\cite{DBLP:conf/mlsys/ChenSSMSMBWR25}, an open-source framework for evaluating autonomous AIOps agents in interactive microservice cloud environments.

Our recovery protocol consists of one initial attempt followed by at most one repair attempt.
Note that this one-repair-attempt design is an evaluation protocol, not a framework limitation.
Each task is first executed without recovery feedback and evaluated by the native evaluator.
Only tasks that remain unresolved after the initial attempt enter the repair phase.
The repair attempt keeps the subject agent, task, tool environment, evaluator, and execution budget fixed, and differs only in the additional recovery information supplied to the agent, thereby isolating the marginal effect of recovery feedback from gains due to repeated retries.

Unless otherwise specified, the agents use Qwen3.5-Plus for task execution, while \name{} and summary-based baselines use Claude Sonnet 4.5 for diagnosis or feedback generation.
Although absolute recovery rates may depend on diagnosis-model capability, using the same feedback model across \name{} and summary-based baselines allows us to compare the structured recovery pipeline against generic trace summarization while controlling for the feedback model.

\subsubsection{Baselines}
\label{sec:eval-baselines}
We compare \name{} with three feedback baselines under the same recovery protocol.
All baselines use the same subject agent, task, evaluator, execution budget, and repair-attempt interface as \name{}.
They differ only in the feedback provided before the repair attempt.

\emph{Outcome Only} provides only the evaluator feedback from the failed initial attempt.
This baseline tests whether the native failure signal, without execution telemetry or diagnosis, is sufficient to guide recovery.
\emph{AgentOps Summary}~\cite{dong2024agentops} and \emph{LangSmith Summary}~\cite{langsmith2026docs} serve as the observability summary baselines.
For each baseline, we collect the execution trace exposed by the corresponding observability system, summarize it with Claude Sonnet 4.5, and combine the summary with evaluator feedback.
These baselines test whether generic summaries of observability traces are sufficient for recovery support.

\emph{\name{} Full Pipeline} is the full-method condition, where the formatted recovery hint block is generated from the telemetry report, structured diagnosis, and \emph{Guidance Gate}.

\subsubsection{Evaluation Metrics}
\label{sec:eval-metrics}
In Table~\ref{tab:main-recovery}, \emph{Initial Unres.} denotes the number of tasks that remain unresolved after the initial attempt, and \emph{Rate} denotes the fraction of those tasks resolved after the repair attempt.
We also report \emph{Top-1}, where a case is counted as correct when the main failure cause reflected in the method's structured diagnosis matches the independently reviewed failure cause.
Top-1 evaluates diagnosis quality rather than end-to-end task success.

To quantify runtime overhead, we use three metrics.
\emph{Telemetry volume} is measured by the number of recorded spans and tool calls.
\emph{Prompt overhead} is measured by the number of additional prompt tokens introduced by the recovery information supplied to the subsequent attempt.
\emph{Feedback-generation latency} is measured from the end of the failed initial attempt to the completion of telemetry report construction, structured diagnosis, \emph{Guidance Gate} processing, and recovery-hint formatting.

\subsection{RQ1: End-to-End Recovery Effectiveness}
\label{sec:eval-recovery}
Table~\ref{tab:main-recovery} reports diagnosis quality and recovery performance under the same one-repair-attempt protocol.
Across 257 initially unresolved tasks, \name{} achieves 65.37\% Top-1 diagnosis accuracy and recovers 56 cases, corresponding to a 21.79\% recovery rate.
Compared with the strongest non-\name{} baselines, \name{} improves Top-1 diagnosis accuracy by 43.58 percentage points and recovery rate by 12.45 percentage points.
The larger improvement in diagnosis accuracy than in recovery rate suggests a diagnosis--recovery gap: accurate failure-cause identification is necessary but not sufficient for successful recovery under fixed environment and budget constraints.

The advantage of \name{} is consistent across settings.
On SWE-bench, \name{} achieves 83.33\% Top-1 accuracy and 28.43\% recovery, exceeding the best non-\name{} baseline by 53.92 and 13.72 percentage points, respectively.
On EnterpriseOps-Gym, \name{} reaches 57.55\% Top-1 accuracy and 11.32\% recovery, improving over the strongest non-\name{} baseline by 53.78 and 5.66 percentage points.
On AIOpsLab, \name{} matches the strongest non-\name{} baseline on Top-1 accuracy at 44.90\%, but achieves a higher recovery rate of 30.61\% versus 16.33\%.

\emph{Outcome Only} recovers only 6 of the 257 initially unresolved tasks, showing that final evaluator feedback alone is a weak basis for recovery.
The observability-summary baselines recover 24 tasks overall, but their Top-1 accuracy remains much lower than \name{}, at 14.79\% and 21.79\%.
This suggests that generic trace summaries provide useful runtime context, but often fail to preserve the signal-specific anchors needed to identify the most recovery-relevant failure cause and produce bounded recovery guidance.
The AIOpsLab result further clarifies the distinction between diagnosis and recovery.
Although \name{} and \emph{LangSmith Summary} achieve the same Top-1 score, \name{} recovers substantially more cases.
In interactive service environments, recovery requires operationally specific guidance, such as which configuration to inspect, which check to rerun, which verification signal to wait for, and which premature action to avoid.

Overall, RQ1 indicates that \name{} improves both diagnostic quality and downstream recovery rate, while also revealing a gap between diagnosis and recovery.

\subsection{RQ2: Actionable and Gated Recovery Guidance}
\label{sec:eval-failure-evidence}


To examine the diagnosis--recovery gap, RQ2 studies whether \name{} can convert failed-run evidence into executable and bounded recovery guidance through the \emph{Guidance Gate}.
We first present a representative case study, and then analyze category-wise diagnosis alignment and recovery outcomes.

\subsubsection{Case Study: From Diagnosis to Bounded Guidance}
\label{sec:eval-case-study}

\begin{figure}[t]
    \centering
    \includegraphics[width=0.8\columnwidth]{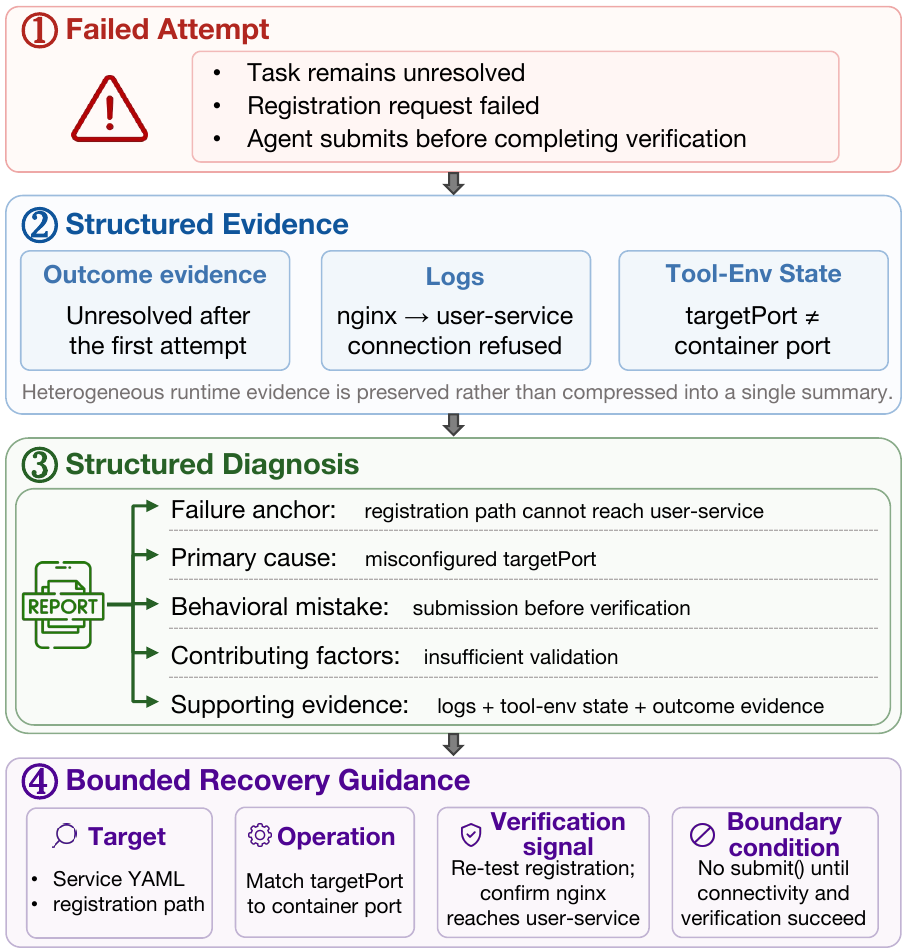}
    \caption{AIOpsLab case illustrating how \name{} preserves failed-run evidence, derives a structured diagnosis, and produces bounded recovery guidance for the subsequent attempt.}
    \label{fig:case-study}
\end{figure}

Figure~\ref{fig:case-study} shows a representative AIOpsLab case.
The initial attempt remains unresolved because the registration path cannot reach \texttt{user-service}, and the agent submits before completing verification.
\name{} preserves three recovery-critical evidence sources: outcome evidence showing that the task remains unresolved, logs showing an \texttt{nginx} connection failure to \texttt{user-service}, and tool--environment state showing a mismatch between the Service \texttt{targetPort} and the actual container port.

The \emph{Diagnosis Layer} converts this evidence into a structured diagnosis: the failure anchor is that the registration path cannot reach \texttt{user-service}, the primary cause is a misconfigured \texttt{targetPort}, and the behavioral mistake is premature submission before verification.
This diagnosis explains the failed run, but it does not by itself specify how the subsequent attempt should proceed.
The \emph{Guidance Gate} turns the diagnosis into bounded recovery guidance by applying the Grounding Check and Actionability Filter, including intervention-scope validation.
In this case, the guidance can be expressed with four actionability fields:
\begin{itemize}[leftmargin=*]
    \item \emph{Target:} the Service YAML and the registration path involving \texttt{user-service}.
    \item \emph{Operation:} compare \texttt{targetPort} with the actual container port and correct the mismatch if present.
    \item \emph{Verification signal:} re-test the registration request and confirm that \texttt{nginx} can reach \texttt{user-service}.
    \item \emph{Boundary condition:} do not call \texttt{submit()} until connectivity and registration verification both succeed.
\end{itemize}
This case illustrates the value of the \emph{Guidance Gate}: diagnosis explains the failure, whereas bounded guidance turns the diagnosis into an executable instruction for the subsequent attempt.

\begin{table}[t]
\centering
\small
\setlength{\tabcolsep}{3.0pt}
\caption{Category-wise diagnosis alignment and recovery under the \name{} full pipeline.
\emph{Aligned} is the fraction of diagnoses matching the independently reviewed failure cause; 
\emph{Recovered} is the fraction of all initially unresolved cases in the category that are solved in the subsequent attempt.}
\label{tab:rq2-hint-hit}
\begin{tabular*}{\columnwidth}{@{}
p{0.43\columnwidth}
>{\centering\arraybackslash}m{0.08\columnwidth}
>{\centering\arraybackslash}m{0.22\columnwidth}
>{\centering\arraybackslash}m{0.22\columnwidth}
@{}}
\toprule
Reviewed failure cause & Cases & Aligned & Recovered \\
\midrule
Insufficient validation & 83 & 81 (97.59\%) & 21 (25.30\%) \\
Tool/subprocess failure handling & 47 & 20 (42.55\%) & 3 (6.38\%) \\
State/workflow error & 42 & 35 (83.33\%) & 6 (14.29\%) \\
Patch/submission workflow & 35 & 12 (34.29\%) & 15 (42.86\%) \\
Retry/no-progress loop & 26 & 6 (23.08\%) & 5 (19.23\%) \\
Runtime/environment handling & 24 & 14 (58.33\%) & 6 (25.00\%) \\
\midrule
\textbf{Total} & 257 & 168 (65.37\%) & 56 (21.79\%) \\
\bottomrule
\end{tabular*}
\end{table}

\subsubsection{Category-wise Results}
\label{sec:eval-rq2-category}

Table~\ref{tab:rq2-hint-hit} reports diagnosis alignment and recovery by reviewed failure category under the \name{} full pipeline.
Overall, \name{} aligns with the independently reviewed failure cause in 168 of 257 cases, but only 56 cases are recovered, confirming that correct diagnosis is important but not sufficient for executable recovery.

The category-level results further clarify this diagnosis--recovery gap.
\emph{Insufficient validation} and \emph{state/workflow error} show strong diagnosis alignment, reaching 97.59\% and 83.33\%, respectively, because they often expose explicit anchors such as failing checks, premature submission, missing state transitions, or incorrect entity linkage.
However, \emph{state/workflow error} reaches only 14.29\% recovery, suggesting that recognizing an incorrect workflow state is easier than repairing it under the same execution budget.
Recovery is highest for \emph{patch/submission workflow}, suggesting that localized submission-stage corrections are often executable once surfaced in the feedback.
By contrast, \emph{tool/subprocess failure handling} has the lowest recovery rate at 6.38\%.
This does not mean such failures are hard to observe; rather, their evidence can be explicit, but successful recovery often requires strategy adaptation beyond a localized correction.
For \emph{runtime/environment handling}, some cases involve recoverable configuration or dependency issues, while others are dominated by runtime conditions outside the agent's direct control.
Accordingly, the \emph{Guidance Gate} bounds the feedback around verifiable agent-side actions when possible, and around verification, evidence collection, or avoiding premature completion when direct repair is not supported.

In summary, RQ2 shows that \name{} goes beyond generic summarization by converting failed-run evidence into structured diagnosis and evidence-grounded recovery feedback, while the remaining diagnosis--recovery gap reflects that successful repair still depends on whether the required recovery path is executable by the agent and environment.

\subsection{RQ3: Practical Overhead}
\label{sec:eval-overhead}
To evaluate the practical overhead of \name{}, we measure telemetry volume, prompt overhead, and feedback-generation latency.
Under \name{} instrumentation, the median failed run contains 578 spans and 164 tool calls on SWE-bench, 46 spans and 17 tool calls on EnterpriseOps-Gym, and 135 spans and 64 tool calls on AIOpsLab.
These differences reflect task structure: repository-level repair requires repeated file inspection, editing, and test execution, whereas enterprise workflow and service-operation tasks typically involve shorter environment-interaction sequences.

The final recovery hint block remains compact across all three settings.
On average, \name{} adds 1.07K prompt tokens on SWE-bench, 0.45K on EnterpriseOps-Gym, and 0.39K on AIOpsLab.
These values remain within the same order of magnitude as the observability-summary baselines, indicating that \name{} does not require substantially larger feedback prompts.
This result is consistent with the design of \name{}: the telemetry report preserves rich failed-run evidence, while structured diagnosis, \emph{Guidance Gate} processing, and hint formatting select only recovery-relevant content for the subsequent attempt.

Across the 257 analyzed cases, the median feedback-generation latency is 210.43 seconds, with setting-level medians of 190.45 seconds on SWE-bench, 300.68 seconds on EnterpriseOps-Gym, and 130.50 seconds on AIOpsLab.
This latency is incurred once after the failed initial attempt and before the repair attempt starts.
It covers telemetry report construction, structured diagnosis, \emph{Guidance Gate} processing, and recovery-hint formatting, and does not modify the subject agent's execution budget.

Taken together, these results indicate that \name{} scales as a side-channel recovery layer, with moderate telemetry volume, compact prompt overhead, and bounded feedback-generation latency.

%% file: sections/deployment.tex
\section{Deployment-Oriented Prototype for IcM Service Diagnosis}
\label{sec:deployment-icm}

In Microsoft Incident Management (IcM), service incidents are 
investigated by diagnosis agents that retrieve heterogeneous operational evidence, including monitor alerts, service metadata, dependency graphs, and historical incident records, and then propose mitigation actions through a structured workflow~\citep{DBLP:conf/icse/AhmedGBZZR23,DBLP:conf/sigsoft/GoelHSGPBZR24,DBLP:conf/icse/ShettyBKRN021,chen2024rcacopilot,jiang2024xpert,zhang2024gpt4rca}.
When such a diagnosis-agent execution fails to resolve an incident, on-call engineers often inspect traces, review retrieved evidence, adjust prompts, and rerun the agent manually.
This process is time-consuming and does not systematically reuse the runtime evidence already produced by the failed attempt.

To validate \name{}’s industrial integration boundary, we build a deployment-oriented proof-of-concept integration. 
It defines the input--output contract and examines whether \name{} can be attached to an existing diagnosis-agent workflow without changing the agent
policy, toolset, evaluator, prompt budget, or execution logic.
In this section, we focus on the engineering interface, integration boundary, and lessons for industrial adoption.

\subsection{Prototype Scope and Input--Output Contract}
\label{sec:deployment-icm-io}

The prototype targets the diagnosis-agent execution loop rather than the full incident-management lifecycle.
It does not replace incident triage, mitigation execution, escalation, or human approval.
Instead, \name{} observes a diagnosis-agent run as a side channel and produces recovery artifacts when the run fails or remains unresolved.

Table~\ref{tab:icm-io-contract} summarizes the input--output contract of the prototype.
The key design point is non-intrusive integration: the diagnosis agent keeps the same model, tools, evaluator, prompt budget, and execution logic, while \name{} records boundary events and produces artifacts after the run.

\begin{table*}[t]
\centering
\small
\renewcommand{\arraystretch}{1.05}
\setlength{\tabcolsep}{4pt}
\caption{Input--output contract and prototype interfaces for integrating \name{} into an IcM diagnosis-agent workflow.}
\label{tab:icm-io-contract}
\begin{tabular*}{\textwidth}{@{\extracolsep{\fill}}
p{0.08\textwidth}
p{0.21\textwidth}
p{0.23\textwidth}
p{0.28\textwidth}
p{0.13\textwidth}
@{}}
\toprule
Stage & Prototype interface & Input to \name{} & Output from \name{} & Agent impact \\
\midrule
Run start &
Session initialization &
Incident context, task, tools, telemetry links, model ID &
Run metadata and tool snapshot &
None \\
\addlinespace[1pt]

During run &
Boundary-event recording &
Model/tool calls, retrieved evidence, claims, checks, exceptions &
Telemetry spans, correlation IDs, evidence refs., outcome observations &
Side-channel only \\
\addlinespace[1pt]

Run end &
Session finalization and report generation &
Final status, evaluator outcome, run metadata &
Telemetry report with evidence, diagnosis, and Guidance Gate output &
None \\
\addlinespace[1pt]

Handoff &
Report export and hint formatting &
Telemetry report and incident context &
Incident artifact or recovery hint block &
No policy/tool/evaluator/budget change \\
\bottomrule
\end{tabular*}
\end{table*}


The four interfaces capture common execution boundaries in IcM diagnosis-agent workflows: run initialization, boundary-event recording, report finalization, and handoff.
More generally, they apply to tool-using or task-evaluable agents whose executions expose comparable boundary events, such as agent calls, tool interactions, validation results, and final outcomes.

\subsection{Integration Design}
\label{sec:deployment-icm-prototype}

To keep the integration lightweight, the prototype exposes a small set of interface-level adapters rather than requiring framework-internal hooks.
A service team only needs to wrap the existing diagnosis-agent loop with a telemetry session and register boundary callbacks for model calls, tool calls, validation results, and final outcomes.
In practice, this requires only run-level session creation, boundary-event recording, and report finalization, rather than changes to the agent's prompt, tool schema, or orchestration logic.

The integration follows a fail-open design.
If \name{} is unavailable, the diagnosis agent continues with its original workflow.
If the execution raises an exception, the telemetry session is finalized in the cleanup path when possible, so the failed run can still produce an execution record.
This property is important for IcM-style workflows, where recovery support must not introduce a new operational failure mode.

The prototype supports two handoff modes.
In a human-facing workflow, the telemetry report is attached to the incident record as a diagnosis artifact, helping on-call engineers review the evidence, failure anchor, and candidate verification or corrective action.
In an iterative-agent workflow, the formatted recovery hint block is supplied to the subsequent attempt as compact feedback.
Both modes consume the same report artifact.

The Guidance Gate is deterministic by design. 
To make the guidance decision transparent, reproducible, and auditable, it applies the \emph{Grounding Check} and \emph{Actionability Filter}, including intervention-scope validation, before allowing structured diagnosis to contribute diagnosis-derived recovery guidance.

In our controlled evaluation, the median feedback-generation latency is 210.43 seconds across 257 failed runs.
For human-facing workflows, the report can be generated off the critical path; for automated reruns, this latency is treated as feedback-generation time before the subsequent attempt, not as part of the agent's execution budget.
This separation supports incremental industrial adoption by allowing teams to use \name{} first as an auditable diagnosis artifact and later as automated recovery feedback.

\subsection{Lessons Learned}
\label{sec:deployment-lessons}

\textbf{Lesson 1: Recovery support needs boundary evidence, not only final outcomes.}
The prototype showed that useful recovery artifacts require events from multiple boundaries: model calls, tool calls, tool returns, validation attempts, exceptions, and evaluator outcomes.
Final success or failure labels are insufficient because they do not explain where the agent lost progress or which action should be revisited.

\textbf{Lesson 2: Recovery artifacts should serve both human review and agent retry.}
The same telemetry report should serve two consumers.
For on-call engineers, it should be readable as an incident artifact that identifies the failure anchor, primary cause, and supporting evidence.
For iterative agents, it should be formatted into a compact recovery hint block with verifiable actions and stopping conditions.


\textbf{Lesson 3: Adoption depends on preserving the existing execution boundary.}
The prototype suggests that recovery support should be integrated at workflow boundaries rather than inside the agent's trusted execution path.
Accordingly, \name{} should remain a wrapper or boundary observer rather than an inline dependency.
This preserves the agent's policy, tools, evaluator, prompt budget, and execution logic in IcM-style workflows, while avoiding a new blocking component in the incident-diagnosis path.

\textbf{Lesson 4: Live deployment should evaluate operational usefulness, not only diagnosis accuracy.}
The proof-of-concept validates the integration boundary and artifact flow, but not production impact on live incidents.
A full IcM deployment should measure whether the report reduces engineer inspection effort, whether generated guidance is accepted by on-call engineers, and whether the side-channel integration remains robust under production load.
These metrics would assess the operational usefulness of the produced artifacts, beyond validating artifact generation and handoff.

%% file: sections/discussion.tex
\section{Discussion and Limitations}
\label{sec:discussion}

Our evaluation has two main limitations.
First, the one-repair-attempt protocol measures single-step recoverability rather than the practical ceiling of iterative recovery.
Multi-round protocols would clarify whether \name{} continues to provide marginal gains across debugging loops or whether recovery saturates after the first guided attempt.
Second, the dependence of recovery on diagnosis-model capability remains to be quantified.
Same-model and model-swap ablations, such as using the task-execution model for diagnosis generation, would further separate the contribution of the structured recovery pipeline from that of the diagnosis model.
Future evaluations will incorporate these extensions to better characterize the robustness and generality of \name{} under iterative recovery settings and across diagnosis models with different capabilities.

%% file: sections/conclusion.tex
We presented \name{}, a failure-anchored framework that converts heterogeneous runtime telemetry into structured diagnosis and bounded recovery guidance for software engineering agents through three coupled stages: a Telemetry Layer, a Diagnosis Layer, and a Guidance Gate.
Evaluation on 257 initially unresolved cases across three settings shows 65.37\% Top-1 diagnosis accuracy and a 21.79\% recovery rate, exceeding the strongest non-\name{} baseline by 43.58 and 12.45 percentage points, respectively.
The results confirm that accurate diagnosis is necessary but insufficient without grounded, bounded, and executable guidance.
A lightweight prototype integration for Microsoft IcM service diagnosis further demonstrates that \name{} can be attached to existing diagnosis-agent workflows without modifying the agent policy, toolset, or execution budget, and that the engineering principles behind its design, including fail-open instrumentation, deterministic guidance gating, and boundary-level adaptation, are applicable to broader industrial software engineering agent workflows.

%% file: sections/data_availablity.tex
We have made our source code and datasets publicly available through an anonymous repository at \url{https://anonymous.4open.science/r/6114a1c7/README.md}.